\documentclass[%
 reprint,
 amsmath,amssymb,
 aps,
]{revtex4-2}

\usepackage{graphicx}
\usepackage{dcolumn}
\usepackage{bm}

\begin{document}

\preprint{APS/123-QED}

\title{Measurement and completely decoherence}

\author{Weijing Li}
\email{liweijing20@bupt.edu.cn}
\affiliation{%
School of Science,Beijing University of Posts and Telecommunications, People’s Republic of China
}%

\date{\today}

\begin{abstract}
By virtue of quantum coherence resource measure, we show that the dephasing measurement on a coherence basis can transfer the coherence contained in system into environment totally, which gives a quantification of decoherence.
\end{abstract}

\maketitle


\section{Introduction}
Quantum decoherence is the loss of quantum coherence. As long as there exists a definite phase relation between different quantum states of a quantum system, the system is said to be coherent. Decoherence was first introduced in 1970 by the German physicist H. Dieter Zeh \cite{zeh1970interpretation} and has been a subject of active research since the 1980s \cite{schlosshauer2005decoherence}. Decoherence has been developed into a complete framework and can be
viewed as the loss of information from a system into the environment, since every system is loosely coupled with the energetic state of its surroundings.

Quantum coherence underlies phenomena such as quantum interference and multipartite entanglement that play
a central role in the applications of quantum physics and quantum information science. Despite the fundamental importance of quantum coherence, the development of a rigorous theory of quantum coherence as a physical resource has been initiated only recently. Indeed, T. Baumgratz, M. Cramer, M. B. Plenio introduced a rigorous framework for the quantification of coherence and identify intuitive and easily computable measures of coherence in 2014 \cite{baumgratz2014quantifying}, and since then plethora of coherence measures had been formulated. Each of these measures captures some physical traits of quantum coherence with specific operational interpretations, see Ref. \cite{streltsov2017colloquium}for a comprehensive review.

Decoherence is intimately related to quantum measurement process. Given a quantum system, after a measurement, the essential quantum superposition property disappears-decoherence occurs. It is desirable to exploit the quantitative characterization of decoherence process via the coherence measures within the quantum resource framework.

In this note, we show that after a measurement in the coherence basis, the original coherence contained in principal system can be transposed into the environmental completely, which echos the issue of decoherence.

\section{Main results}
Given a quantum state $\rho$ who lives in a finite-dimensional Hilbert space $\mathcal{H}$, which can be spanned by the coherence basis
$\mathcal{H}=span\{|i\rangle\}$. The quantum state $\rho$ can be written as
\begin{equation*}
    \rho=\sum_{ij}\langle i|\rho|j\rangle|i\rangle\langle j|.
\end{equation*}

Define the total dephasing operation $\Phi$ by, which completely destroys coherence of states,
\begin{align*}
    \Phi(\rho)=\sum_i \langle i|\rho|i\rangle|i\rangle\langle i|.
\end{align*}

After performing a total dephasing operation $\Phi$, the post-measurement state can be represented as
\begin{equation*}
    \sigma=\Phi(\rho)=\sum_i p_i |i\rangle\langle i|,
\end{equation*}
where $p_i= \langle i|\rho|i\rangle$.

Since the purification of a quantum state contains all of the information encoded in quantum state, it is natural to consider the purification of $\rho$,
\begin{equation*}
    |\Psi\rangle=\sqrt{\rho}\otimes I |\Sigma\rangle,
\end{equation*}
where $|\Sigma\rangle=\sum_i|ii\rangle$ is the maximally entangled state, up to a constant.

For pure state $|\Psi\rangle$, after performing the total dephasing operation on principal system, the post-measurement state reduces to
\begin{align*}
    \sum_i |i\rangle\langle i|\otimes I  |\Psi\rangle \langle\Psi|i\rangle\langle i|\otimes I.
\end{align*}

The probability of outcome $i$ can be written as 
\begin{align*}
    q_i=&\langle\Psi| i\rangle\langle i|\otimes I|\Psi\rangle\\
       =& \sum_j |\langle i|\sqrt{\rho}|j\rangle|^2\\
       =&\sum_j \langle i|\sqrt{\rho}|j\rangle\langle j|\sqrt{\rho}|i\rangle\\
       =&\langle i|\rho|i\rangle=p_i.
\end{align*}
Since after measurement the state for principal system contains zero coherence, it is desirable to consider the reduced environment state. After tracing out the principal system, the reduced environment state is
\begin{align*}
    \tau=&\text{Tr}_\mathcal{H}\{\sum_i |i\rangle\langle i|\otimes I  |\Psi\rangle \langle\Psi|i\rangle\langle i|\otimes I\}\\
    =&\sum_{ij}\langle j|\rho|i\rangle|i\rangle\langle j|.
\end{align*}
One can see that $\tau$ is just the entry-wise complex conjugate of $\rho$ in coherence basis representation. It is directly to observe that for any coherence measure involved the the modular of the entries of
quantum state, $\rho$ and $\tau$ have equal coherence. For example, the $l_1$ norm coherence measure:
\begin{equation*}
    C_{l_1}(\rho)=\sum_{i\ne j}|\rho_{ij}|,
\end{equation*}
we have 
\begin{equation}
     C_{l_1}(\rho)=C_{l_1}(\tau).
\end{equation}

This interesting fact can be interpreted as that the measurement process transfers the coherence contained in original quantum state into the environment completely, and $l_1$ norm coherence measure can be used to quantify both coherence and decoherence.

Besides of $l_1$ norm coherence measure $C_{l_1}(\rho)$, other typical coherence measures, relative entropy for example, are also of same property. 

Recall that the quantum relative entropy is written as
\begin{align*}
    S(\rho||\sigma)=\text{Tr}\left(\rho \log \rho\right)- \text{Tr}\left(\rho \log \sigma\right)
\end{align*}
and the relative entropy coherence measure is defined as 
\begin{align*}
    C_R(\rho)=\min_{\sigma \in \mathcal{I}} S(\rho||\sigma),
\end{align*}
where $\mathcal{I}$ is the set of incoherent states, that is, all incoherent sates $\sigma \in \mathcal{I}$ are of the form $\sigma=\sum_i p_i |i\rangle\langle i|$ with probabilities $p_i$. Relative entropy coherence measure $C_R(\rho)$ has the operational interpretation of distillable coherence and admits a closed expression
\begin{align*}
    C_R(\rho)=S(\Phi(\rho))-S(\rho)
\end{align*}
in terms of the von Neumann entropy
\begin{align*}
   S(\rho)=-\text{Tr}\left(\rho \log \rho\right).
\end{align*}

One can observe that 
\begin{align}
    C_R(\rho)=C_R(\tau).
\end{align}
In fact, for the spectral decomposition of $\rho=\sum_k \lambda_k |\psi_k\rangle\langle \psi_k|$, it is easy to verify that $\tau$ admits a spectral decomposition
\begin{align*}
    \tau=\sum_k \lambda_k |\psi_k^*\rangle\langle \psi_k*|
\end{align*}
where $|\psi_k^*\rangle$ is the entry-wise complex-conjugate of $|\psi_k\rangle$. Due to the positivity of $\rho$, it must hold that 
\begin{align*}
    S(\Phi(\rho))=S(\Phi(\tau)) \quad\text{and}\quad  S(\rho)=S(\tau),
\end{align*}
Hence we obtain
\begin{align*}
     C_R(\rho)=C_R(\tau).
\end{align*}

Another interesting example is the coherence measure induced by Wigner-Yanase skew information $C_{WY}(\rho)$ \cite{yu2017quantum}. Note that the Wigner-Yanase skew information is defined as \cite{wigner1963information}
\begin{align*}
    I(\rho,H)=\text{Tr}(\rho H^2)-\text{Tr}(\sqrt{\rho}H\sqrt{\rho}H)
\end{align*}
where $H=H^{\dagger}$ plays the role of conversed quantity. $C_{WY}(\rho)$ is constructed by
\begin{align*}
    C_{WY}(\rho)=&\sum_i I(\rho, |i\rangle \langle i|)\\
                =&1- \sum_i \langle i|\sqrt{\rho}|i\rangle^2.
\end{align*}
By direct calculation it is easy to find that
\begin{align}
    C_{WY}(\rho)=C_{WY}(\tau).
\end{align}
Indeed, one only need to verify
\begin{align*}
     &\langle i|\sqrt{\rho}|i\rangle\\
    =&\sum_k \sqrt{\lambda_k} |\langle i|\psi_k\rangle|^2\\
    =&\sum_k \sqrt{\lambda_k} |\langle i|\psi_k^*\rangle|^2\\
    =&\langle i|\sqrt{\tau}|i\rangle.
\end{align*}

In the contrast, robustness of coherence introduced in \cite{napoli2016robustness} seems challenging to verify equation relations similar to equations (1-3), though it admit some different operational interpretations. Nevertheless, by focusing on the specific dephasing operation, we can use coherence measures to quantify quantum decoherence quantitatively.

\section{Discussions}
Quantum decoherence has been a subject of active research over forty years and various methods had been proposed to interpret its mechanisms quantitatively, for example, einselection \cite{zurek2003decoherence} and semigroup approach \cite{bauer2012iterated}. In this note we quantify the decoherence process by virtue of the quantum coherence resource measures quantitatively. Our results is in agreement with decoherence issue and quantum coherence resource theory.

\bibliography{apssamp}

\end{document}